\documentclass[12pt]{article}
\usepackage{a4wide}
\usepackage{amssymb, amsmath}
\usepackage{xcolor}
\usepackage{hyperref}
\usepackage[normalem]{ulem}
\begin{document}
{\renewcommand{\thefootnote}{\fnsymbol{footnote}}

\begin{center}
{\LARGE Effective line elements and black-hole models\\ in canonical (loop)
  quantum gravity} \\
\vspace{1.5em}
Martin Bojowald$^1$\footnote{e-mail address: {\tt bojowald@gravity.psu.edu}},  Suddhasattwa Brahma$^{2}$\footnote{e-mail address: {\tt suddhasattwa.brahma@gmail.com}} and Dong-han Yeom$^{2,3}$\footnote{e-mail address: {\tt innocent.yeom@gmail.com}}
\\
\vspace{0.5em}
$^1$ Institute for Gravitation and the Cosmos,\\
The Pennsylvania State
University,\\
104 Davey Lab, University Park, PA 16802, USA\\
\vspace{0.5em}
$^2$ Asia Pacific Center for Theoretical Physics,\\
Pohang 37673, Korea\\
$^3$ Department of Physics, POSTECH,\\
Pohang 37673, Korea\\
\vspace{1.5em}
\end{center}
}

\setcounter{footnote}{0}

\newcommand{\bea}{\begin{eqnarray}}
\newcommand{\eea}{\end{eqnarray}}
\renewcommand{\d}{{\mathrm{d}}}
\renewcommand{\[}{\left[}
\renewcommand{\]}{\right]}
\renewcommand{\(}{\left(}
\renewcommand{\)}{\right)}
\newcommand{\nn}{\nonumber}

\def\be{\begin{equation}}
\def\ee{\end{equation}}

\def\SU{\text{SU}}
\def\su{\mathfrak{su}}
\newcommand{\cC}{{\mathcal C}}
\newcommand{\cG}{{\mathcal G}}
\newcommand{\cD}{{\mathcal D}}
\newcommand{\cH}{{\mathcal H}}
\newcommand{\cL}{{\mathcal L}}
\newcommand{\cM}{{\mathcal M}}

\newcommand{\Su}{\color{red}}

\begin{abstract}
	Canonical quantization is often used to suggest new effects in
	quantum gravity, in the dynamics as well as the structure of
	space-time. Usually, possible phenomena are first seen in a modified
	version of the classical dynamics, for instance in an effective
	Friedmann equation, but there should also be implications for a
	modified space-time structure. Quantum space-time effects, however,
	are often ignored in this setting because they are not obvious: they
	require a careful analysis of gauge transformations and the anomaly
	problem. It is shown here how modified space-time structures and
	effective line elements can be derived unambiguously, provided an
	off-shell anomaly-free system of modified constraints exists. The
	resulting effective line elements reveal signature change as an
	inescapable consequence of non-classical gauge transformations in
	the presence of holonomy modifications. The general framework is
	then specialized to black-hole models in loop quantum gravity. In
	contrast to previous studies, a self-consistent space-time structure
	is taken into account, leading to a new picture of black-hole
	interiors.
\end{abstract}

\section{Introduction}
The covariance problem is highly non-trivial in modified or quantum canonical
gravity. Not just the dynamics but also the structure of space-time may
change, and the correct space-time structure is to be derived from the
theory. A self-consistent space-time structure within a given theory is
important not only for an interpretation of the theory but also for explicit
derivations based on effective line elements. A line element ${\rm d}s^2$, by
definition, provides a distance measure independent of the choice of
coordinates. In any theory with classical covariance, based on Riemannian
geometry, the metric tensor $g_{\mu\nu}$ has the correct transformation
properties such that the familiar equation
\begin{equation}
	{\rm d}s^2=g_{\mu\nu}{\rm d}x^{\mu}{\rm d}x^{\nu}
\end{equation}
results in a coordinate-independent expression. 

In canonical gravity, the components of $g_{\mu\nu}$ are split into different
types of variables: the spatial metric $q_{ab}$, the shift vector $N^a$, and
the lapse function $N$. They now define the line element in the form
\cite{ADM}
\begin{equation} \label{dsCan}
	{\rm d}s^2 = -N^2{\rm d}t^2+ q_{ab}\left({\rm d}x^a+N^a{\rm d}t\right)
	\left({\rm d}x^b+N^b{\rm d}t\right)
\end{equation}
with a time coordinate $t$ and spatial coordinates $x^a$, $a=1,2,3$ in
$x^{\mu}=(t,x^a)$. There is a one-to-one correspondence between the components
of $g_{\mu\nu}$ and the components of $q_{ab}$ together with $N^a$ and
$N$. However, in the canonical formulation, the coefficients in the space-time
line element play rather different roles: The spatial metric $q_{ab}$ provides
the phase-space coordinates, along with momenta $p^{cd}$ related to extrinsic
curvature. Lapse and shift, on the other hand, are multipliers of first-class
constraints $H[N]$ and $D[N^a]$, and therefore determine a gauge but do not
have non-zero momenta.

Gauge transformations, generated by $H[N]$ and $D[N^a]$ via Poisson brackets,
seem to change only phase-space variables, that is, the spatial components
$q_{ab}$ of the metric (\ref{dsCan}). A generic coordinate transformation, on
the other hand, should change all the coefficients in the space-time line
element. There should therefore be gauge transformations of lapse and shift as
well, in addition to those of $q_{ab}$. In classical canonical gravity, it is
known that such transformations are indeed implied by the constrained system
given by $H[N]$ and $D[N^a]$, in particular by the consistent interplay of
equations of motion and gauge transformations generated by the
constraints. (See \cite{CUP} for details.) Such an interplay relies on the
{\em off-shell} form of Poisson brackets of the constraints, or the
hypersurface-deformation brackets
\begin{eqnarray}
	\{D[N_1^a],D[N_2^b]\} &=& D[{\cal L}_{N_1}N_2^a] \label{DD} \\
	\{H[N],D[N_1^a]\} &=& -H[{\cal L}_{N_1}N] \label{HD} \\
	\{H[N_1],H[N_2]\} &=& \pm D[q^{ab}(N_1\partial_b N_2-N_2\partial_b N_1)] \label{HH}
\end{eqnarray}
where the $\pm$ in the last equation indicates space-time signature.  (An
equivalent formulation \cite{LapseGauge} may use an extended phase space in
which lapse and shift do have momenta, subject to additional constraints that
they vanish. These additional constraints are not relevant for space-time
transformations and therefore do not affect considerations of space-time
structure.)

However, it seems that these general properties have not always been taken
into account in modified or quantum canonical gravity. In this setting,
motivated by different kinds of formal aspects of quantization, one modifies
some of the constraints, in most cases the Hamiltonian constraint $H[N]$. A
large class of examples includes homogeneous models of quantum cosmology, in
which only a single constraint, $H[N]$ with spatially constant $N$, is
non-trivial and implies a modified Friedmann equation when quantized. The
modified Friedmann equation then leads to modified evolution equations for the
scale factor $a(t)$, which may show interesting new effects. One could then
insert a modified scale factor in an ``effective'' line element ${\rm
	d}s^2=-N(t)^2{\rm d}t^2+a(t)^2{\rm d}s_k^2$, with the appropriate spatially
isotropic line element ${\rm d}s_k^2$ and a lapse function $N(t)$
corresponding to one's choice of time. But without an analysis of space-time
structure, such a line element is coordinate dependent and therefore
meaningless if $H[N]$ is modified such that the constraints no longer obey the
classical hypersurface-deformation brackets. In such a case, gauge
transformations of $a$ according to modified constraints are not consistent
with coordinate transformations of $t$ and $x^a$. In some cases, the
constraint brackets may still be closed but modified, in which case a suitable
effective line element requires additional modifications. And if the
constraint brackets no longer close, gauge transformations are broken and no
effective line element exists. These statements underline the importance of
the off-shell brackets of constraints, the derivation of which presents one of
the most challenging tasks in approaches to canonical quantum gravity.

Fortunately, there are some models of canonical quantum gravity in which
closed modified constraint brackets can be derived in their off-shell
form. These include some versions of cosmological perturbations
\cite{ConstraintAlgebra,ScalarHol,ScalarHolInv}, and spherically symmetric
models with non-perturbative inhomogeneity
\cite{JR,LTBII,ModCollapse,HigherSpatial}. Here, we will use the latter models
because their equations are less lengthy and show the relevant features more
clearly. After a review of covariance in classical canonical gravity, we will
show how all the required ingredients of covariance can be derived in modified
canonical models, provided one has access to the off-shell brackets of
constraints. As an application, we will obtain effective line elements and
black-hole models with crucial new ingredients not considered before.

In particular, we shall show that given a deformation of the bracket between
two modified Hamiltonian constraints, the same deformation function appears in
the time-time component of the effective line-element. If the bracket is
deformed such that its structure function changes sign, the time-time
component of the effective line element turns positive, while the remaining
contributions remain positive definite. This result unequivocally demonstrates
the change in signature of the space-time metric which had earlier been
deduced more indirectly from the change in sign of the
hypersurface-deformation brackets, or from field equations they imply
\cite{Action,SigChange}. Thus, signature-change is a necessary effect of
having deformed hypersurface-deformations such that their structure function
changes sign, as long as one is careful in deriving an invariant line-element
that is consistent with the modified gauge transformations of the effective
theory.

\section{Effective equations in spherically symmetric models of loop quantum gravity}

In spherically symmetric models, the canonical line element (\ref{dsCan})
takes the form
\begin{equation}
	{\rm d}s^2 = -N(t,x)^2{\rm d}t^2+ q_{xx}(t,x) \left({\rm d}x+N^x(t,x) {\rm
		d}t\right)^2+ q_{\varphi\varphi}(t,x) {\rm d}\Omega^2
\end{equation}
with two independent metric components $q_{xx}$ and $q_{\varphi\varphi}$, and
just one non-zero component $N^x$ of the shift vector field, all depending on
$t$ and just the radial spatial coordinate $x$. 

In order to be closer to models of loop quantum gravity, we express the metric
components in terms of a densitized spatial triad with components $E^x$ and
$E^{\varphi}$, such that
\begin{equation}
	q_{xx} = \frac{(E^{\varphi})^2}{E^x} \quad,\quad q_{\varphi\varphi} = E^x\,.
\end{equation}
For our purposes, it is sufficient to view these equations as part of a
canonical transformation. (For details of the derivation of these and some of
the following relations, see \cite{SymmRed,SphSymm,SphSymmHam}. In general,
$E^x$ could take positive or negative values, indicating the orientation of
space. We will assume $E^x>0$ without loss of generality.) The triad
components then acquire momenta $K_x$ and $K_{\varphi}$ such that
\begin{equation}
	\{K_x(x),E^x(y)\} = 2G\delta(x,y) \quad\mbox{and}\quad
	\{K_{\varphi}(x),E^{\varphi}(y)\} = G\delta(x,y)\,.
\end{equation}

The relationship between the momenta and partial derivatives of triad
variables (or extrinsic curvature) is determined by the constraints, solving
$\dot{E}^x=\{E^x,H[N]+D[N^a]\}$ and
$\dot{E}^{\varphi}=\{E^{\varphi},H[N]+D[N^a]\}$ for $K_x$ and
$K_{\varphi}$. In classical gravity, the constraints are 
\begin{eqnarray}
	H[N] &=&-\frac{1}{G} \int {\rm d} x\, N \left( \frac{E^{\phi}}{2 \sqrt{E^x}}
	K_{\varphi}^2 +  K_{\varphi} \sqrt{E^{x}} K_x + \frac{E^{\phi}}{2
		\sqrt{E^x}}  - \frac{((E^x)')^2}{8 \sqrt{E^{x}} E^{\phi}} \right.\nonumber\\ 
	& & \hspace{2cm} \left. + \frac{\sqrt{E^x} (E^x)' (E^{\phi})'}{2
		(E^{\phi})^2} - \frac{\sqrt{E^x}(E^x)''}{2E^{\phi}}\right)
\end{eqnarray}
and
\begin{equation}
	D[N^x] = \frac{1}{2G} \int {\rm d}x\, N^x \left(2K_\varphi^\prime\, E^\varphi - K_x
	(E^x)^\prime \right)\,.
\end{equation}
Their off-shell brackets mimick the full ones, with 
\begin{equation}
	q^{xx}=\frac{E^x}{(E^{\varphi})^2}
\end{equation}
replacing $q^{ab}$ in (\ref{HH}).

\subsection{Holonomy-modified effective constraints}

Holonomy modifications in spherically symmetric models change the dependence
of the Hamiltonian constraint on $K_{\varphi}$. Similarly to the
full theory of loop quantum gravity \cite{LoopRep,ALMMT}, spherically
symmetric models in a loop quantization have operators only for exponentials
$\exp(i\delta K_{\varphi})$ of $K_{\varphi}$, which is part of the
components of a U(1)-connection \cite{SphSymm}. Prior to quantization, the
Hamiltonian constraint therefore is modified by correction terms with higher
powers of $K_{\varphi}$, such that the classical polynomial form is extended
to a periodic function \cite{SphSymmHam}, again mimicking the construction
in the full theory \cite{QSDI}. The diffeomorphism constraint, by contrast,
is represented directly via its finite spatial transformations, without
modifications. In order to analyze potential consequences of loop
quantizations, one can therefore consider holonomy modifications in the
Hamiltonian constraint, with the important condition that they not break
gauge transformations. This condition is implemented by making sure that the
constraints remain first class, that is, their Poisson brackets are still
closed after holonomy modification.

The constraint brackets remain closed if we use a Hamiltonian constraint of
the form
\begin{eqnarray}
	H[N] &=&- \frac{1}{G} \int \text{d}x\, N \left( \frac{E^{\phi}}{2 \sqrt{E^x}}
	f_1(K_{\varphi}) +  f_2(K_{\varphi}) \sqrt{E^{x}} K_x + \frac{E^{\phi}}{2
		\sqrt{E^x}}  - \frac{((E^x)')^2}{8 \sqrt{E^{x}} E^{\phi}} \right.\nonumber\\ 
	& & \hspace{2cm} \left. + \frac{\sqrt{E^x} (E^x)' (E^{\phi})'}{2
		(E^{\phi})^2} - \frac{\sqrt{E^x}(E^x)''}{2E^{\phi}}\right)
\end{eqnarray}
with two functions $f_1(K_{\varphi})$ and $f_2(K_{\varphi})$, such that
\cite{JR,HigherSpatial}
\begin{equation} \label{f1f2}
	f_2(K_{\varphi})= \frac{1}{2}\frac{{\rm d}f_1(K_{\varphi})}{{\rm
			d}K_{\varphi}}\,.
\end{equation}
The structure function in the bracket of two modified Hamiltonian constraints
then acquires a factor of
\begin{equation}\label{beta}
	\beta(K_{\varphi})=\frac{1}{2}\frac{{\rm d}^2f_1(K_{\varphi})}{{\rm
			d}K_{\varphi}^2}\,.
\end{equation}

In these derivations, the diffeomorphism constraint is left unmodified: In
most constructions of the kinematics of loop quantum gravity, one does not use
holonomies to quantize the diffeomorphism constraint but rather represents its
finite action \cite{ALMMT}. This procedure does not suggest holonomy
modifications in the diffeomorphism constraint. Moreover, in the Hamiltonian
constraint, only the $K_{\varphi}$-dependence is modified while the
$K_x$-dependence remains linear. This form is motivated by differences in the
roles played by these two extrinsic-curvature components in holonomies. While
$K_x$ appears in extended, non-local holonomies, exponentiating the radially
integrated $K_x$, $K_{\varphi}$ appears in ``point'' holonomies
\cite{FermionHiggs} which just exponentiate a multiple of $iK_{\varphi}$ at a
given point $x$. (See \cite{SphSymm} for the corresponding quantum
representation.) Holonomy effects on $K_x$ can therefore be made small by
choosing short curves for holonomies, while $K_{\varphi}$-modifications remain
unchanged. The same arguments suggest that any non-linear modifications of the
$K_x$-dependence should also introduce higher spatial derivatives of $K_x$
from a derivative expansion of extended holonomies. However, no anomaly-free
set of modified constraints is known in this form \cite{HigherSpatial}.

The function $\beta(K_{\varphi})$ is close to one only when $f_1(K_{\varphi})$
is close to the classical behavior, $f_1(K_{\varphi})\approx K_{\varphi}^2$,
which is usually the case for small $K_{\varphi}$. For other values, the
constraint brackets do not have the classical form, and therefore the
coefficients of a canonical line element (\ref{dsCan}) no longer transform in
the classical way consistent with coordinate transformations. In such regimes,
we have a modified space-time structure which requires a careful analysis
before a coordinate-independent effective line element can be defined. We will
pursue this question in the next section, but first indicate how strongly the
space-time structure can be modified: Near a local maximum of $f_1$, $\beta$
is negative, and the structure function in the bracket of two Hamiltonian
constraints changes sign, indicating signature change according to
(\ref{HH}). An analysis taking this new effect into account will, in general,
lead to drastically different results, compared with one that assumes a
Lorentzian ``effective'' line element in which one just inserts metric
components subject to a modified dynamics. (See for instance \cite{Loss}.)

The remaining function, $f_1(K_{\varphi})$, is not restricted by the condition
that constraint brackets close.  It is usually chosen such as to mimic
holonomy components, for instance
\begin{equation} \label{f1}
	f_1(K_{\varphi})=\frac{\sin^2(\delta K_{\varphi})}{\delta^2}
\end{equation}
with a real parameter $\delta$. In this case,
\begin{equation}
	f_2(K_{\varphi})=\frac{\sin(2\delta K_{\varphi})}{2\delta}
\end{equation}
and
\begin{equation}
	\beta(K_{\varphi}) = \cos(2\delta K_{\varphi})\,.
\end{equation} 
The function (\ref{f1}) is chosen as a simple example of a
function that corresponds to a U(1)-holonomy in spherically symmetric
models (replacing SU(2)-holonomies in the full theory). 
The precise choice depends on quantization ambiguities \cite{HigherSpinLQC}, as  almost any
periodic function could be obtained from U(1) characters.

\subsection{Evolution, gauge and coordinates}

The functions $f_1$ and $f_2$ appear in modified field equations generated by
$H[N]+D[N^a]$: 
\begin{equation}
	\dot{F} = \{F,H[N]+D[N^a]\}
\end{equation}
for any phase-space function $F$. For the basic variables, we obtain
\begin{eqnarray}
	\dot{E}^x &=&   2 N \sqrt{E^x} \; f_2(K_{\varphi}) + N^x (E^x)'\label{Exdot} \\
	\dot{E}^{\varphi} &=&   N \sqrt{E^x} K_x \frac{{\rm d}f_2(K_{\varphi})}{{\rm
			d}K_{\varphi}} +  \frac{N E^{\varphi}}{2\sqrt{E^x}} \frac{{\rm
			d}f_1(K_{\varphi})}{{\rm d}K_{\varphi}} + (N^x E^{\varphi})'\label{Ephidot}
\end{eqnarray}
and
\begin{eqnarray}
	\dot{K}_x & =&  -N\left(-\frac{E^{\varphi}}{2(E^x)^{3/2}}
	f_1(K_{\varphi}) + \frac{1}{\sqrt{E^x}} f_2(K_{\varphi}) K_x-
	\frac{E^{\varphi}}{2(E^x)^{3/2}}   + \frac{((E^x)')^2}{8 E^{\varphi} (E^x)^{3/2}} +
	\frac{(E^x)'(E^{\varphi})'}{2\sqrt{E^x}(E^{\varphi})^2}\right.\label{Kxdot}\nonumber\\
	&&\left.-
	\frac{(E^x)''}{2\sqrt{E^x}E^{\varphi}}\right)
	-\left(\frac{N(E^x)'}{2\sqrt{E^x}E^{\varphi}}\right)'+
	\left(\frac{N\sqrt{E^x}(E^{\varphi})'}{(E^{\varphi})^2}\right)'
	+\left(\frac{N\sqrt{E^x}}{E^{\varphi}}\right)'' +(N^xK_x)'\\
	\dot{K}_{\varphi} & = &  N\left( -\frac{1}{2 \sqrt{E^x}} (1 + f_1(K_{\varphi})) 
	-\frac{((E^x)')^2}{8\sqrt{E^x}(E^{\varphi})^2}+
	\frac{\sqrt{E^x}(E^x)'(E^{\varphi})'}{(E^{\varphi})^3}-
	\frac{\sqrt{E^x}(E^x)''}{2(E^{\varphi})^2}\right)\nonumber\\
	&&+
	\left(\frac{N\sqrt{E^x}(E^x)'}{2(E^{\varphi})^2}\right)'
	+N^xK_{\varphi}'\,. \label{Kphidot} 
\end{eqnarray}

For a given choice of $N$ and $N^x$, solutions for the phase-space variables
are obtained in a fixed gauge. Because the modified constraints remain first
class, they still generate gauge transformations
\begin{equation}
	\delta F = \{F,H[\epsilon^0]+D[\epsilon^x]\}
\end{equation}
consistent with the dynamics: The evolution of gauge-transformed initial
values is equal to a gauge transformation of evolved initial values. For the
phase-space variables, gauge transformations 
\begin{eqnarray}
	\delta E^x &=&   2 \epsilon^0 \sqrt{E^x} \; f_2(K_{\varphi}) + \epsilon^x
	(E^x)'\label{deltaEx} \\ 
	\delta E^{\varphi} &=&   \epsilon^0 \sqrt{E^x} K_x \frac{{\rm
			d}f_2(K_{\varphi})}{{\rm 
			d}K_{\varphi}} +  \frac{\epsilon^0 E^{\varphi}}{2\sqrt{E^x}} \frac{{\rm
			d}f_1(K_{\varphi})}{{\rm d}K_{\varphi}} + (\epsilon^x
	E^{\varphi})'\label{deltaEphi} 
\end{eqnarray}
and
\begin{eqnarray}
	\delta K_x & =&  -\epsilon^0\left(-\frac{E^{\varphi}}{2(E^x)^{3/2}}
	f_1(K_{\varphi}) + \frac{1}{\sqrt{E^x}} f_2(K_{\varphi}) K_x-
	\frac{E^{\varphi}}{2(E^x)^{3/2}}   + \frac{((E^x)')^2}{8 E^{\varphi} (E^x)^{3/2}} +
	\frac{(E^x)'(E^{\varphi})'}{2\sqrt{E^x}(E^{\varphi})^2}\right.
	\label{deltaKx}\nonumber\\ 
	&&\left.-
	\frac{(E^x)''}{2\sqrt{E^x}E^{\varphi}}\right)
	-\left(\frac{\epsilon^0(E^x)'}{2\sqrt{E^x}E^{\varphi}}\right)'+
	\left(\frac{\epsilon^0\sqrt{E^x}(E^{\varphi})'}{(E^{\varphi})^2}\right)'
	+\left(\frac{\epsilon^0\sqrt{E^x}}{E^{\varphi}}\right)'' +(\epsilon^xK_x)'\\
	\delta K_{\varphi} & = &  \epsilon^0\left( -\frac{1}{2 \sqrt{E^x}} (1 +
	f_1(K_{\varphi}))  -\frac{((E^x)')^2}{8\sqrt{E^x}(E^{\varphi})^2}+
	\frac{\sqrt{E^x}(E^x)'(E^{\varphi})'}{(E^{\varphi})^3}-
	\frac{\sqrt{E^x}(E^x)''}{2(E^{\varphi})^2}\right)\nonumber\\
	&&+
	\left(\frac{\epsilon^0\sqrt{E^x}(E^x)'}{2(E^{\varphi})^2}\right)'
	+\epsilon^xK_{\varphi}' \label{deltaKphi} 
\end{eqnarray}
have a form similar to the
evolution equations because they are generated by the same constraints, but
their meaning is quite different. For effective line elements, it is important
to consider evolution equations as well as gauge transformations, and their
interplay.

We will now approach the question of how modified gauge transformations could
be related to coordinate transformations. The components of the spatial metric
are subject to modified gauge transformations according to (\ref{deltaEx}) and
(\ref{deltaEphi}). Gauge transformations of the remaining components of an
effective line element, $N$ and $N^x$, are not as obvious, but they are
uniquely determined by the constrained system. In the next section we review
these aspects in general form, and then apply them to our modified spherically
symmetric systems.

\section{Canonical covariance}

Covariance is not realized in a manifest way in canonical gravity, but it must
still be present, given that the Hamiltonian formulation is equivalent to the
more common Lagrangian one. Because covariance does not directly meet the eye
in canonical versions of modified or quantum gravity, it is easy to overlook
its importance when one works in a particular gauge without worrying whether
other gauges would produce the same physical effects. If covariance is
implemented properly, on the other hand, consistency conditions are imposed on
possible quantum modifications, eliminating some (but perhaps not all) of the
usual ambiguities. There are therefore important reasons to study covariance
in canonical gravity, which remains one of the major open problems in
approaches such as loop quantum gravity.

In this section, we briefly review the general features of covariance in
classical canonical gravity, and then make them more explicit in generic
spherically symmetric models. These details will allow us to see how
covariance can be implemented in spherically symmetric models which are no
longer classical but modified by holonomy effects motivated by loop quantum
gravity. We will notice an important new term which is implied by holonomy
modifications and necessary to obtain well-defined and gauge invariant
effective line elements. This new term has not been included in any one of the
previously existing black-hole models of loop quantum gravity.

\subsection{Constraints, structure functions, and line elements}

In canonical gravity, space-time coordinate transformations are replaced by
gauge transformations of phase-space functions $F(q_{ab},p^{cd})$ generated by
the diffeomorphism and Hamiltonian constraints, $D[\epsilon^i]$ and
$H[\epsilon^0]$. The diffeomorphism constraint generates deformations within a
spatial slice used for the canonical decomposition, while the Hamiltonian
constraint generates deformations of a spatial slice along its normal
direction $n^a$ in space-time. A combination of both transformations,
generated by $H[\epsilon^0]+D[\epsilon^i]$, therefore implies a deformation
along a space-time vector field $\epsilon^a= \epsilon^0 n^a+\epsilon^i s_i^a$,
using a basis $s_i^a$, $i=1,2,3$ of the spatial tangent bundle. For a generic
phase-space function $F$, we obtain the gauge transformation
\begin{equation}
	\delta F = \{F,H[\epsilon^0]+D[\epsilon^i]\}\,.
\end{equation}

In order to compare these space-time deformations with infinitesimal
coordinate changes, one should first translate the tangential-normal
decomposition of a space-time vector field into a space-time decomposition,
with reference to the time-evolution vector field $t^a=Nn^a+N^a$ of an ADM
formulation \cite{ADM}. Here, $N$ and $N^a$ are the lapse function and shift
vector field of the space-time metric which is being transformed. A space-time
vector field $\xi^a$, referring to the time direction and the previous spatial
basis $s_i^a$, is decomposed as $\xi^a = \xi^0 t^a+ \xi^i s_i^a$. If we want
to find a gauge transformation which implements a coordinate transformation
along $\xi^a$, we should choose the components $\epsilon^0$ and $\epsilon^i$
such that $\epsilon^a$ and $\xi^a$ are the same vector field, just written in
different bases (see also \cite{PhaseSpaceCoord}):
\begin{equation}
	\epsilon^0 n^a+\epsilon^i s_i^a = \frac{\epsilon^0}{N} t^a+
	\left(\epsilon^i-\frac{N^i}{N} \epsilon^0\right) s_i^a = \xi^0 t^a+\xi^i
	s_i^a\,.
\end{equation}
Comparing coefficients, a gauge transformation by $(\epsilon^0,\epsilon^i)$
should therefore correspond to a coordinate transformation in a direction
given by
\begin{equation} \label{xieps}
	\xi^0 = \frac{\epsilon^0}{N} \quad,\quad \xi^i = \epsilon^i-\frac{N^i}{N}
	\epsilon^0\,.
\end{equation}
We then have the identity
\begin{equation} \label{Gauge}
	{\cal L}_{\xi} F = \{F,H[\epsilon^0]+D[\epsilon^i]\}
\end{equation}
for phase-space functions $F(q_{ab},p^{cd})$, provided the identities
(\ref{xieps}) hold for the components of $\xi^a$ and $\epsilon^a$.

As mentioned briefly in the introduction, in canonical gravity, gauge
transformations generated by $H$ and $D$ act only on the spatial metric
$q_{ab}$ and its momentum $p^{cd}$. These constraints do not directly generate
transformations of lapse and shift, which form the time-time and time-space
components of the space-time line element
\begin{equation}
	{\rm d}s^2 = -N^2{\rm d}t^2+ q_{ab} ({\rm d}x^a+N^a{\rm d}t) ({\rm
		d}x^b+N^b{\rm d}t)\,.
\end{equation}
However, a generic coordinate transformation clearly changes not just $q_{ab}$
but also the components $g_{0a}$ of the space-time metric. In order to discuss
covariance and line elements in canonical gravity, one could consider the
extended phase space in which also lapse and shift have momenta, that is, the
phase space on which primary constraints have not yet been solved
\cite{LapseGauge}. Alternatively, the transformations follow from the
requirement that the canonical equations of motion are gauge covariant
\cite{ScalarGaugeInv,CUP}.

The latter viewpoint is instructive because it makes it clear that the
off-shell constraint brackets, forming Dirac's hypersurface-deformation
algebroid, are important in this context: Canonical equations of motion for
the phase-space variables, formulated in a specific gauge, depend on the
corresponding lapse and shift. A gauge transformation (\ref{Gauge}) changes
the phase-space variables, and the transformed variables will obey equations
of motion consistent with the original ones only if lapse and shift are
properly transformed too. We are looking for a commutation property: The
evolution of gauge-transformed initial data must be equal to a gauge
transformation of the evolved initial data. Since equations of motion as well
as gauge transformations are generated by the same constraints $H$ and $D$,
the former by $H[N]+D[N^a]$ and the latter by $H[\epsilon]+D[\epsilon^a]$, the
commutation of gauge and evolution depends on the Poisson brackets of the
constraints. An explicit calculation \cite{CUP} shows that a constrained system
with structure functions $F_{AB}^D$, such that the constraints obey
\begin{equation}
	\{C_A,C_B\}= F_{AB}^D C_D\,,
\end{equation}
leads to covariant equations of motion generated by $C[N^A]$, provided
the multipliers $N^A$ are subject to a gauge transformation
\begin{equation} \label{deltaNF}
	\delta_{\epsilon}N^A= \dot{\epsilon}^A+ N^B\epsilon^C F_{BC}^A\,.
\end{equation}
(Summed-over indices such as $D$ may be continuous, implying
integrations.)

For canonical gravity, the structure functions can be read off from the
hypersurface-deformation brackets (\ref{DD})--(\ref{HH}). They imply that the
lapse function $N$ should be subject to a gauge transformation
\begin{equation} \label{LapseGauge}
	\delta_{\epsilon}N= \dot{\epsilon}^0+\epsilon^i\partial_iN- N^i\partial_i
	\epsilon^0
\end{equation}
and the shift vector field to a gauge transformation
\begin{equation} \label{ShiftGauge}
	\delta_{\epsilon}N^i = \dot{\epsilon}^i+\epsilon^j\partial_jN^i-
	N^j\partial_j\epsilon^i- q^{ij}(N\partial_j\epsilon^0-
	\epsilon^0\partial_jN)\,.
\end{equation}
If these transformations are combined with a gauge transformation of the
spatial metric, the full space-time metric is transformed according to a
space-time coordinate change.

The explicit result shows that the off-shell brackets and their structure
functions are relevant for the correct transformation of lapse and shift. This
result underlines the importance of the full anomaly problem of modified or
quantum gravity, making sure that the modified constraints obey not just some
closed or partially closed system, but a modified version of the
hypersurface-deformation brackets. In general, it is not enough to have
closure on a phase-space on which some of the constraints have been solved (as
in \cite{AnoFree}), or of a reformulated constraint system (such as partial
Abelianizations in \cite{LoopSchwarz}). Only if one has access to the full
off-shell brackets, in such a way that the usual hypersurface-deformation
brackets are obtained in the classical limit, can one realize modified
versions of the transformations (\ref{LapseGauge}) and (\ref{ShiftGauge}), a
crucial ingredient of effective line elements. This conclusion will be made
much more explicit by our analysis of spherically symmetric modls in what
follows. (Examples of systems which formally have closed brackets of modified
constraints but are not covariant exist, see \cite{SphSymmCov,GowdyCov}.)

\subsection{Spherically symmetric line elements}

By definition, the variables $E^x$ and $E^{\varphi}$ used in spherically
symmetric models are such that the spatial line element is given by
\begin{equation}
	{\rm d}s^2 = \frac{(E^{\varphi})^2}{E^x} {\rm d}x^2+ E^x ({\rm
		d}\vartheta^2+\sin^2\vartheta{\rm d}\varphi^2)\,.
\end{equation}
Therefore, the spatial metric has components
\begin{equation}
	q_{xx} = \frac{(E^{\varphi})^2}{E^x} \quad\mbox{and}\quad
	q_{\varphi\varphi} = E^x\,.
\end{equation}
A corresponding space-time line element is
\begin{equation} \label{dsSpaceTime}
	{\rm d}s^2 = -N(x,t)^2{\rm d}t^2+ q_{xx}(x,t) ({\rm d}x+N^x(x,t){\rm d}t)^2+
	q_{\varphi\varphi}(x,t) ({\rm
		d}\vartheta^2+\sin^2\vartheta{\rm d}\varphi^2)\,.
\end{equation}

We can now apply a spherically symmetric coordinate transformation along the
vector field $(\xi^0,\xi^x)=(\epsilon^0/N, \epsilon^x-(N^x/N)\epsilon^0)$,
using (\ref{xieps}), directly to the line element by inserting $t'=t+\xi^0$
and $x'=x+\xi^x$ for $t$ and $x$. The calculations are lengthy, but not as
long as without symmetry assumptions, and they will ultimately be instructive
when we discuss modified constraints.

After inserting transformed coordinates $x^a+\xi^a$ in the line element, we
collect coefficients of ${\rm d}x^2$, ${\rm d}x{\rm d}t$ and ${\rm d}t^2$ in
the new line element, in this order. The new coefficient of ${\rm d}x^2$
receives a contribution from a first-order expansion of $q_{xx}(x', t')$, one
from expanding
\begin{equation}
	{\rm d}x'={\rm     d}(x+\epsilon^x-(N^x/N)\epsilon^0)= {\rm d}x+
	(\epsilon^x-(N^x/N)\epsilon^0)^{\bullet}{\rm d}t+
	(\epsilon^x-(N^x/N)\epsilon^0)'{\rm d}x
\end{equation}
in ${\rm d}x^2$, as well as one from the old ${\rm d}x{\rm d}t$ term in which
we include, for now, only the ${\rm d}x$-term in
\begin{equation}
	{\rm d}t'={\rm d}(t+\epsilon^0/N)=
	{\rm d}t+(\epsilon^0/N)^{\bullet}{\rm d}t+ (\epsilon^0/N)'{\rm d}x\,.
\end{equation}
The latter two originate from cross-terms in squares, and therefore count
twice. We obtain
\begin{equation} \label{deltaqxx}
	\delta q_{xx}=\frac{\epsilon^0}{N} \dot{q}_{xx}+
	\left(\epsilon^x-\frac{N^x}{N}\epsilon^0\right) q_{xx}'+ 2q_{xx}
	\left(\epsilon^x-\frac{N^x}{N}\epsilon^0\right)' + 2q_{xx}N^x
	\left(\frac{\epsilon^0}{N}\right)'\,.
\end{equation}

Similarly, the new ${\rm d}x{\rm d}t$-term receives several contributions from
which, using $\delta q_{xx}$, we have
\begin{equation} \label{NxSphSymm}
	\delta N^x = \dot{\epsilon}^x+ \epsilon^x (N^x)'- N^x(\epsilon^x)'-
	\frac{1}{q_{xx}} \left(N(\epsilon^0)'-\epsilon^0N'\right)\,.
\end{equation}
Finally, we derive
\begin{equation} \label{NSphSymm}
	\delta N = \dot{\epsilon}^0+ N'\epsilon^x-N^x(\epsilon^0)'
\end{equation}
using both $\delta q_{xx}$ and $\delta N^x$. (See App.~\ref{a:LapseShift} for
details.)  These results confirm the general expressions (\ref{LapseGauge})
and (\ref{ShiftGauge}); note in particular a component of the inverse metric
in (\ref{NxSphSymm}). (There is no inverse $q_{\varphi\varphi}$ because the
shift vector cannot have a $\varphi$-dependent $\varphi$-component if it
preserves spherical symmetry.)

\subsection{Holonomy modifications}

As a result of the preceding calculations, we see that a line element of the
form (\ref{dsSpaceTime}) is meaningful, that is, coordinate independent, only
if lapse and shift are transformed according to (\ref{NSphSymm}) and
(\ref{NxSphSymm}). For classical gravity, these transformations are implied by
the general result (\ref{deltaNF}) for constrained systems. However, in the
presence of holonomy modifications, some of the structure functions are
modified. The same general result, (\ref{deltaNF}), then implies that $\delta
N^x$ has to be modified to
\begin{equation} \label{NxSphSymmbeta}
	\delta N^x = \dot{\epsilon}^x+ \epsilon^x (N^x)'- N^x(\epsilon^x)'-
	\frac{\beta}{q_{xx}} \left(N(\epsilon^0)'-\epsilon^0N'\right)
\end{equation}
where $\beta/q_{xx}$ is the new structure function in the bracket of two
Hamiltonian constraints. 

Similarly, even though this structure function does not appear explicitly in
(\ref{LapseGauge}), the classical (\ref{NxSphSymm}) is used in a derivation of
(\ref{NSphSymm}) for spherically symmetric models, as shown in
App.~\ref{a:LapseShift}. This derivation can no longer proceed in the
classical form if structure functions are modified such that $\beta\not=\pm
1$. In particular, details of the derivation of (\ref{NSphSymm}) show that the
$q_{xx}$-independent contributions to the new ${\rm d}t^2$-term are given by
\begin{equation} \label{NDetail}
	-2N\dot{N}\frac{\epsilon^0}{N}- 2NN'
	\left(\epsilon^x-\frac{N^x}{N}\epsilon^0\right)-
	2N^2\left(\frac{\epsilon^0}{N}\right)^{\bullet} = -2N\dot{\epsilon}^0-
	2NN'\epsilon^x+ 2N^xN'\epsilon^0 \,.
\end{equation}
Upon using $\delta N^x$, a new $q_{xx}$-independent term $2NN^x(\epsilon^0)'-
2N^xN'\epsilon^0$ is generated by the contribution
$q_{xx}^{-1}(N(\epsilon^0)'-\epsilon^0N')$ to $\delta N^x$, and the result
(\ref{NSphSymm}) follows. If $\delta N^x$ is modified to
(\ref{NxSphSymmbeta}), the correct $\delta N$ as required by (\ref{deltaNF})
no longer follows unless further modifications are implemented in just the
right way.

Since the term in $\delta N^x$ relevant for a derivation of $\delta N$ is
multiplied with $\beta$, the classical cancellations leading to
(\ref{NSphSymm}) still happen if all the terms in (\ref{NDetail}) are
multiplied with $\beta$. This is achieved if one performs a coordinate
transformation not of the classical line element (\ref{dsSpaceTime}), but of
an effective line element
\begin{equation} \label{dsEffective} 
	{\rm d}s^2 = -\beta(x,t) N(x,t)^2{\rm d}t^2+
	q_{xx}(x,t) ({\rm d}x+N^x(x,t){\rm d}t)^2+ q_{\varphi\varphi}(x,t) ({\rm
		d}\vartheta^2+\sin^2\vartheta{\rm d}\varphi^2)\,.
\end{equation}
Holonomy modification not only imply corrections to the spatial metric by
modified equations of motion generated by the Hamiltonian constraint, they
also require a new factor of $\beta$ of $N^2$ in the time-time component of
the space-time line element. Signature change for $\beta<0$ is an immediate
consequence.

In the derivation of $\delta N$ from a coordinate transformation we also made
use of $\delta q_{xx}$ in addition to $\delta N^x$. Since the constraints are
modified by holonomy terms, gauge transformations and equations of motion for
$q_{xx}$ are modified as well. However, a direct calculation shows that the
modified equations of motion for $E^x$ and $E^{\varphi}$ imply that the
classical relation (\ref{deltaqxx}) between a gauge transformation of
$q_{xx}=(E^{\varphi})^2/E^x$ and its space and time derivatives remains
unchanged, for any choice of $f_1$ and $f_2$. (See App.~\ref{a:q}.)

The new factor of $\beta$ in the time-time component produces the correct
modification (\ref{NxSphSymmbeta}) as required by the modified constraint
brackets, and it ensures that the derivation of $\delta N$ still results in
(\ref{NSphSymm}), also required by the constraint brackets. The new time-time
component, transformed by a coordinate transformation, now has an expansion
given by
\begin{equation}
	\delta (-\beta N^2) = -2\beta N\delta N- \dot{\beta}
	\epsilon^0 N- 
	\beta' \left(\epsilon^x-\frac{N^x}{N}\epsilon^0\right) N^2
	= -2\beta N \left(\delta N+ \frac{N}{2\beta}\delta \beta\right)
\end{equation}
with 
\begin{equation}\label{absorbBeta}
	\delta N+ \frac{\delta\beta}{2\beta}N= \dot{\epsilon}^0+N'\epsilon^x-
	N^x(\epsilon^0)'+ \frac{1}{2} \frac{\dot{\beta}}{\beta} \epsilon^0+
	\frac{1}{2}\frac{\beta'}{\beta} (N\epsilon^x - N^x\epsilon^0)=
	\frac{(\sqrt{\beta}\epsilon^0)^{\bullet}+ (\sqrt{\beta}N)'\epsilon^x-
		N^x(\sqrt{\beta}\epsilon^0)'}{\sqrt{\beta}}\,,
\end{equation}
assuming $\beta>0$ in the last step, as well as scalar transformation.
Therefore, for $\beta>0$ we have
\begin{equation}
	\delta(-\beta N^2) = -2\sqrt{\beta}N
	\left((\sqrt{\beta}\epsilon^0)^{\bullet}+ (\sqrt{\beta}N)'\epsilon^x-
	N^x(\sqrt{\beta}\epsilon^0)'\right) \,,
\end{equation}
and transformations of the effective line element can be mapped to those of
the classical line element by absorbing a factor of $\sqrt{\beta}$ in the
lapse function $N$ and in the time component $\epsilon^0$ of a gauge
transformation. Alternatively, one can view this as absorbing $\sqrt{\beta}$
in the normal vector $n^a$, which is then no longer a unit vector. The latter
viewpoint is an example for the general result of \cite{Normal} based on
properties of Lie algebroids. If $\beta<0$, $\sqrt{|\beta|}$ can be absorbed
in the same way, provided one maps to a Euclidean 4-dimensional line element.

\section{Specific solutions of effective line elements}

The equations of motion (\ref{Exdot})--(\ref{Kphidot}) form a set of coupled,
non-linear, partial differential equations. An important class of
classical solutions which are easier to derive are stationary ones, which we
can try to look for also in the modified case.  In Eq.~(\ref{Exdot}), the
left-hand side must then be zero, as must be the shift vector: We are looking
for the zeros of one of the holonomy-modification functions,
$f_2(K_\varphi)=0$. Potential candidates of stationary solutions are therefore
such that
\begin{equation}
	K_{\varphi} = 0\quad,\quad K_{\varphi}=\frac{\pi}{2\delta}\quad,\quad \ldots
\end{equation}
(assuming a modification function of the form (\ref{f1}).)
The value of $K_\varphi=0$ corresponds to the classical solution, as shall be
demonstrated in more detail in the next section. The value of $K_\varphi=
\pi/(2\delta)$ may be considered a stationary solution in the deep quantum
regime of strongly modified constraints, corresponding to a Euclidean phase in
an effective line element. The presence of Euclidean solutions means that we
have to generalize the meaning of ``stationary.'' If there are four spatial
dimensions but no time, the intuitive understanding of stationary solutions
does not apply. The formal definition as solutions with a timelike Killing
vector can, however, be generalized to a definition in which any solution with
a Killing vector transversal to the hypersurfaces of a canonical $3+1$
decomposition is considered stationary (or static if the Killing vector is
normal to the hypersurfaces).

Another space(-time) property requires more care. It is not always obvious how
new stationary solutions for large $K_{\varphi}$ should be interpreted. We can
see this difficulty if we first describe what happens in the well-understood
case of the classical Schwarzschild solution in canonical form.  We know that
in this case a classical stationary solution of the canonical equations of
motion corresponds to the entire space-time region outside the horizon. In
this region, $K_{\varphi}=0$ does not identify a unique spatial slice but
rather specifies a gauge condition on the slicing of space-time. In other
regions of the same space-time, however, an equation such as $K_{\varphi}=0$
can have a different meaning. For instance, in the interior, formulated as a
Kantowski-Sachs model, $K_{\varphi}$ is no longer identically zero and a fixed
value then corresponds to a unique spatial hypersurface within a homogeneous
slicing.  The existence of different possible meanings of $K_{\varphi}=0$
depending on the position in space-time and its gauge shows that, without a
further detailed analysis of slicings and gauges, we cannot uniquely decide
which version new stationary solutions with $\delta K_{\varphi}=\pi/2$ should
correspond to.

Additional solutions, with even larger values of $K_\varphi=n\pi/(2\delta)$
for integer $n$, could indicate the existence of further alternating cycles of
Euclidean and Lorentzian phases. However, before embarking on a detailed
analysis of the quantum space(-time) dynamics, it is mute to speculate about
potential transitions or phase-changes between the two classical
signatures. Here, we are mainly interested in developing an understanding of
the approach to larger values of $K_{\varphi}$ and eventually the first
Euclidean region.

\subsection{The classical Schwarzschild solution}

We begin with a simpler task: recovering the classical solution in an
effective line element.  In order to solve the full set of spherically
symmetric equations, we first choose the gauge $E^x = x^2$ and then examine
its flow as generated by the diffeomorphism constraint:
\begin{eqnarray}
	\delta E^x= \{E^x(x), D[N^x]\} = N^x(x) \left(E^x(x)\right)^\prime \approx  2
	N^x x
\end{eqnarray}
vanishes, and therefore the gauge condition is conserved, provided $N^x =
0$. With this first gauge choice, $E^x =x^2$, we solve the diffeomorphism
constraint $D[N^x]=0$ and obtain
\begin{equation}
	K_x = \frac{E^\varphi}{x} K_\varphi^\prime\,.
\end{equation}

We can now rewrite the Hamiltonian constraint with this partial gauge fixing 
\begin{equation}\label{RedHam}
	H[N] = -\frac{1}{2G}\int {\rm d}x\, N \(\(1 + f_1\(K_\varphi \)
	\)\frac{E^\varphi}{x} +  
	2E^\varphi f_2\(K_\varphi\) K_\varphi' -\frac{3x}{E^\varphi} + \frac{2x^2
		\(E^\varphi\)^\prime}{\(E^\varphi\)^2}\)\,.
\end{equation}
It generates a flow given by
\begin{eqnarray}
	\delta_N E^\varphi &:=& \{E^\varphi, H[N]\} = \(\frac{N E^\varphi}{x} -
	\frac{\d}{\d x}\(N E^\varphi\)\) f_2\(K_\varphi\)+ NE^{\varphi} K_{\varphi}'
	\frac{{\rm d}f_2}{{\rm d}K_{\varphi}} \label{EFlowRedHam} \,,\\ 
	\delta_N K_\varphi &:=& \{K_\varphi, H[N]\} =
	-\frac{N}{2x}\(1+f_1\(K_\varphi\)\) - \frac{N}{2}\frac{\d f_1}{\d
		x} +\frac{N x}{2 (E^\varphi)^2} + \frac{N^\prime
		x^2}{\(E^\varphi\)^2}\label{KFlowRedHam} \,. 
\end{eqnarray}
For a stationary solution, $E^{\varphi}$ is conserved. Since $K_{\varphi}$
is constant, $\delta_N K_{\varphi}$ implies an equation for $N$ using
(\ref{KFlowRedHam}), depending on $E^{\varphi}$ restricted by $H[N]=0$ in
(\ref{RedHam}). 

We have already stated that the classical stationary solution should
correspond to $K_\varphi=0$. Since $K_{\varphi}$ is fixed, its flow generated
by $H[N]$, (\ref{KFlowRedHam}), must vanish:
\begin{equation}
	-\frac{N}{2x} +\frac{N x}{2 (E^\varphi)^2} + \frac{N^\prime x^2}{\(E^\varphi\)^2}
	= 0
\end{equation}
or
\begin{equation}
	\(E^\varphi\)^2 = x^2 \(1 + 2x\frac{N^\prime}{N}\)\,.\label{NEphieqn}
\end{equation}
(One may view $K_{\varphi}=0$ as a gauge-fixing condition.)

Finally, the Hamiltonian constraint (\ref{RedHam}) implies a
differential equation for $E^\varphi$,
\begin{equation}
	\(E^\varphi\)^3 - 3 x^2 E^\varphi + 2 x^3 \frac{\d E^\varphi}{\d x} =
	0\,, 
\end{equation}
with the general solution
\begin{equation}\label{ClassEphieqn} 
	E^\varphi = \frac{x}{\sqrt{1 + c/x}}\,.
\end{equation}
The integration constant $c = -2M$ can be fixed, as usual, by referring
to the mass for large $x$. Using this solution in (\ref{NEphieqn}) then
implies a differential equation
\begin{equation}\label{ClassNeqn} 
	2 \frac{\d   N}{\d x} = -\frac{Nc}{x(x+c)}
\end{equation}
for the lapse function, with the solution $N(x) = \sqrt{1 + c/x}$. (An
additional integration constant in this solution would simply rescale the
lapse function.)

We can now formulate a line element. Since $\beta(K_{\varphi})=1$ for
$K_{\varphi}=0$, the effective line element, obtained by inserting our
solutions in 
\begin{equation}
	{\rm d} s^2 = -\beta N^2 \d t^2 + \frac{\(E^\varphi\)^2}{E^x} \left({\rm d}
	x + N^x{\rm d} t\right)^2 + E^x {\rm d}\Omega^2\,,
\end{equation}
is identical with the classical Schwarzschild line element
\begin{equation} \label{Schwarzschild}
	{\rm d} s^2 = -\left(1 - \frac{2M}{x}\right){\rm d} t^2 + \frac{1}{1 -
		2M/x} {\rm d} x^2 +x^2{\rm d}\Omega^2\,.
\end{equation}

\subsection{Holonomy-modified interior}

The same arguments as used in classical general relativity show that the
values of $x<2M$ can no longer correspond to a stationary solution because,
according to (\ref{Schwarzschild}), $x$ then plays the role of a time
coordinate and the metric depends on $x$. The gauge choice $K_{\varphi}=0$ is
therefore unavailable in the interior. Moreover, $E^x=x^2$ is not a good gauge
choice if $x$ becomes a time coordinate (unless one would like to use
$\sqrt{E^x}$ as internal time). 

The interior, classically, is no longer stationary but spatially homogeneous
because the metric does not depend on the new spatial coordinate $t$, and it
remains spherically symmetric. The homogeneity condition $(E^x)'=0$ can be
used as a new gauge choice, also in the modified constraints. (Here and in
what follows, a prime always denotes a spatial derivative, while a dot is a
time derivative. In the interior of (\ref{Schwarzschild}), a prime therefore
refers to a derivative by $t$, while a dot refers to a derivative by $x$. This
convention is used here because it is consistent with the dots and primes in
the general equations of motion (\ref{Exdot})--(\ref{Kphidot}).) The
diffeomorphism constraint then implies $K_{\varphi}'=0$ for non-degenerate
triads. The Hamiltonian constraint, with spatially constant $E^x$ and
$K_{\varphi}$, implies that $K_x/E^{\varphi}$ is spatially constant. And with
this condition $\dot{E}^{\varphi}/E^{\varphi}$, according to (\ref{Ephidot}),
is homogeneous if we assume homogeneous $N$ and $N^x=0$. If we choose
homogeneous initial data, they remain homogeneous. Spatially homogeneous
solutions therefore exist for any choice of holonomy-modification
functions. Moreover, any such solution is connected to the stationary exterior
through a horizon: The classical homogeneous interior, with
$N(T)=1/\sqrt{2M/T-1}$, $E^x(T)=T^2$ and $E^{\varphi}(T)=T\sqrt{2M/T-1}$ where
$T=x$ in terms of the coordinates used in (\ref{Schwarzschild}), has a
vanishing $K_{\varphi}(T)=\dot{E}^x/(2N\sqrt{E^x})=\sqrt{2M/T-1}=0$ at the
horizon. Therefore, holonomy modifications are expected to be small around the
horizon, and the classical matching surface is unmodified.

By these arguments, we may consider modified homogeneous solutions as a
black-hole interior.  Since the classical relationship $T=x$ with an exterior
coordinate may no longer be realized, we call the new time coordinate in the
interior $\eta$. (We will also choose a different lapse function for our
modified interior solutions.)  The function $\sqrt{E^x(\eta)}$ would then be
one of the scale factors of a modified Kantowski--Sachs model, accompanied by
$E^{\varphi}(\eta)/\sqrt{E^x(\eta)}$ as an independent one in an anisotropic
homogeneous model. Their time dependence is given by the previous equations of
motion, (\ref{Exdot})--(\ref{Kphidot}), setting all spatial derivatives equal
to zero and using a spatially constant lapse function. The Hamiltonian
constraint, also with zero spatial derivatives, then implies a Friedmann-type
equation.

It is convenient to choose the lapse function to be $N=\sqrt{E^x}$, an
anisotropic version of conformal time. (For solutions in a different choice,
see \cite{DefSchwarzschild}.) In the equations of motion
\begin{eqnarray}\label{InteriorEOM}
	\dot{E}^x &=& 2 E^xf_2(K_{\varphi})\\
	\dot{E}^{\varphi} &=& E^x K_x \frac{{\rm d}f_2}{{\rm d}K_{\varphi}}+
	\frac{1}{2} E^{\varphi} \frac{{\rm d}f_1}{{\rm d}K_{\varphi}}\\
	&=& E^x K_x \frac{{\rm d}f_2}{{\rm d}K_{\varphi}}+ E^{\varphi}
	f_2(K_{\varphi})\\
	\dot{K}_x &=& \frac{E^{\varphi}}{2E^x} f_1(K_{\varphi})- K_x
	f_2(K_{\varphi})+ \frac{E^{\varphi}}{2E^x}\\
	\dot{K}_{\varphi} &=& -\frac{1}{2}(1+f_1(K_{\varphi}))\,,
\end{eqnarray}
the last one decouples from the rest. For $f_1$ as in (\ref{f1}), it can be
solved by
\begin{equation} \label{Kphieta}
	K_{\varphi}(\eta) = \frac{1}{\delta}
	\arctan\left(\frac{\delta}{\sqrt{1+\delta^2}}
	\tan\left(-\frac{1}{2}\sqrt{1+\delta^2} (\eta-\eta_0)\right)\right)\,.
\end{equation}
(The form of this function resembles generalizations of Chebyshev polynomials
considered in \cite{GenCheb}, but we are not aware of a simplified form in our
case.)

This function is of interest because it appears in the new coefficient $\beta
N^2$ in an effective line element, derived from the deformation function
$\beta(K_{\varphi})=\cos(2\delta K_{\varphi})$ in the bracket of two
spherically symmetric Hamiltonian constraints. In particular, signature change
is indicated by ${\rm sgn}\beta$. Inserting our solution
$K_{\varphi}(\eta)$, we obtain
\begin{equation}
	\beta(\eta)= \frac{1-\frac{\delta^2}{1+\delta^2}
		\tan^2(-\frac{1}{2}\sqrt{1+\delta^2}(\eta-\eta_0))}{1+
		\frac{\delta^2}{1+\delta^2} 
		\tan^2(-\frac{1}{2}\sqrt{1+\delta^2}(\eta-\eta_0))}\,.
\end{equation}
Signature change happens when $\eta$ reaches a value such that
\begin{equation}
	\frac{\delta}{\sqrt{1+\delta^2}}
	\tan\left(-\frac{1}{2}\sqrt{1+\delta^2}(\eta-\eta_0)\right)=1\,.
\end{equation}
``After'' this time, $\eta$ is no longer a time coordinate but can still be
used as the fourth coordinate in the Euclidean region. Changing $\eta$ is then
no longer time evolution but can still be interpreted as a transition between
different slices in the Euclidean core, determined by the homogeneous
gauge. For increasing $\eta$, the value $\beta(\eta)=-1$ of Euclidean space is
approached for $\tan(-\frac{1}{2}\sqrt{1+\delta^2}(\eta-\eta_0))\to\infty$, or
\begin{equation}
	\eta=\eta_0-\frac{\pi}{\sqrt{1+\delta^2}}\,.
\end{equation}
At this point, $\delta K_{\varphi}=\frac{1}{2}\pi$, a value which will be
discussed in more detail in the following subsection.

With our $K_{\varphi}(\eta)$, we obtain
\begin{equation}
	E^x(\eta) = \frac{M^2}{(1+\delta^2)^2} \left(1+2\delta^2
	+\cos\left(\sqrt{1+ \delta^2}(\eta - \eta_0) \right)
	\right)^2
\end{equation}
from (\ref{InteriorEOM}).
Using the Hamiltonian constraint, 
\begin{equation}
	\frac{E^{\varphi}}{2\sqrt{E^x}} f_1(K_{\varphi})+ \sqrt{E^x} K_x
	f_2(K_{\varphi}) + \frac{E^{\varphi}}{2\sqrt{E^x}}=0
\end{equation}
for spatially homogeneous variables, we can decouple the remaining variables
$E^{\varphi}$ and $K_x$: We first obtain
\begin{equation}
	\frac{K_x}{E^{\varphi}} = -\,\frac{1+f_1(K_{\varphi})}{2 E^xf_2(K_{\varphi})}
\end{equation}
from the Hamiltonian constraint, and insert this result in 
\begin{equation}
	\frac{\dot{E}^{\varphi}}{E^{\varphi}} = E^x \frac{K_x}{E^{\varphi}}
	\frac{{\rm d}f_2}{{\rm d}K_{\varphi}}+ f_2(K_{\varphi})= -
	\frac{1+f_1(K_{\varphi})}{2f_2(K_{\varphi})} \frac{{\rm d}f_2}{{\rm
			d}K_{\varphi}}+ f_2(K_{\varphi})\,.
\end{equation}
This equation is slightly simpler if we write it directly for the metric
component $q_{xx}=(E^{\varphi})^2/E^x$:
\begin{equation}
	\frac{\dot{q}_{xx}}{q_{xx}} = -\frac{1+f_1(K_{\varphi})}{2f_2(K_{\varphi})}
	\frac{{\rm d}f_2}{{\rm d}K_{\varphi}}\,,
\end{equation}
using the relation between $f_1$ and $f_2$. The solution for $E^{\varphi}$ is 
\begin{equation}
	E^{\varphi}(\eta) = M\sin\left(\sqrt{1+ \delta^2}(\eta-\eta_0)\right)\,,
\end{equation}
which we can finally use to obtain 
\begin{equation}
	K_x(\eta) = \frac{1+\delta^2}{2M} \;
	\frac{1+2\delta^2+\sqrt{1+\delta^2} -
		\left(\sqrt{1+\delta^2} - 1\right)\cos
		\left(\sqrt{1+ \delta^2}(\eta-\eta_0)\right)}{ 
		\left(1+2\delta^2+\cos\left(\sqrt{1+
			\delta^2}(\eta-\eta)\right)\right)^2}\,.
\end{equation}

The usual choice of $f_1$, (\ref{f1}), also allows us to express the
$K$-dependent terms in the Hamiltonian constraint in terms of time derivatives
of triad components, using the equations of motion: Trigonometric identities
result in
\begin{equation}
	\frac{{\rm d}f_2}{{\rm d}K_{\varphi}} = \sqrt{1-4\delta^2f_2(K_{\varphi})^2}=
	\sqrt{1-\delta^2\left(\frac{\dot{E}^x}{E^x}\right)^2}
\end{equation}
such that 
\begin{equation}
	K_x=\frac{\dot{E}^{\varphi}- E^{\varphi} f_2(K_{\varphi})}{2E^x{\rm
			d}f_2/{\rm d}K_{\varphi}}= \frac{\dot{E}^{\varphi}-
		\frac{1}{2}\frac{E^{\varphi}}{E^x}\dot{E}^x}{2E^x
		\sqrt{1-\delta^2(\dot{E}^x/E^x)^2}}\,. 
\end{equation}
Moreover, 
\begin{equation}
	f_1(K_{\varphi}) = \frac{1-{\rm d}f_2/{\rm d}K_{\varphi}}{2\delta^2}=
	\frac{1-\sqrt{1-\delta^2(\dot{E}^x/E^x)^2}}{2\delta^2}\,.
\end{equation}
Eliminating $K$-components in the Hamiltonian constraint, we therefore obtain
the ``effective'' Hamiltonian
\begin{eqnarray}
	H[N] &=&-L_0 \frac{N}{\sqrt{E^x}} \left( E^{\phi}
	\frac{1-\sqrt{1-\delta^2(\dot{E}^x/E^x)^2}}{4\delta^2}
	+ \frac{\left(\dot{E}^{\varphi}-
		\frac{1}{2}\frac{E^{\varphi}}{E^x}\dot{E}^x\right)\dot{E}^x}{2E^x
		\sqrt{1-\delta^2(\dot{E}^x/E^x)^2}}  
	+\frac{1}{2}E^{\phi}\right)\\
	&=& -\frac{L_0N}{2\sqrt{E^x}\sqrt{1-\delta^2(\dot{E}^x/E^x)^2}} \left( 
	\dot{E}^{\varphi} \frac{\dot{E}^x}{E^x} 
	+ \frac{1}{2\delta^2}E^{\phi}\left((1+2\delta^2)
	\sqrt{1-\delta^2(\dot{E}^x/E^x)^2}-1\right)\right) \nonumber
\end{eqnarray}
for a finite region of coordinate length $L_0=\int{\rm d}x$.

\subsection{Effective line element in the deep quantum regime}

The homogeneous interior evolves such that $K_{\varphi}$ grows  as
we move away from the horizon. (Note that $\eta$ decreases from $\eta_0$ at
the horizon to $\eta_0-\pi/\sqrt{1+\delta^2}$ at the transition of signature
change. Similarly, the time coordinate $t$ in a classical interior,
corresponding to $r$ in the usual Schwarzschild line element, decreases from
$t=2M$ at the horizon to $t=0$ at the singularity. Formally, therefore,
$K_{\varphi}(\eta)$ given in (\ref{Kphieta}) is a decreasing function.)  It
eventually reaches a value where $\beta=0$, and the signature changes to
4-dimensional Euclidean space. If we just follow formal solutions in this
regime, $K_{\varphi}$ still grows in the direction normal to hypersurfaces of
the canonical decomposition, which is no longer a time direction. We should
therefore switch to a 4-dimensional boundary-value problem, or a 2-dimensional
one for $t$ and $x$ in the spherically symmetric reduction. There is not much
known about possible boundary conditions in the deep quantum regime, and
therefore the interior solutions becomes uncertain at this point. However, we
use the fact that large $K_{\varphi}$ will be reached as a motivation to take
a closer look at the new stationary solutions realized for
$K_{\varphi}=\pi/(2\delta)$.

We choose to fix the gauge in the same way as before, using $E^x=x^2$, but
assigning the value $\pi/(2\delta)$ to $K_\varphi$. 
(After a transition to Euclidean signature, $E^x=x^2$ is again a
good gauge choice even though we may remain in the interior.)
For this choice,
$f_1(K_\varphi) = 1/\delta^2$ and $f_2(K_\varphi) = 0$. (Note that the
resulting solutions will not have a classical limit because $f_1$ diverges for
$\delta\to0$.)  We then proceed as before:
Compatibility of the flow of $K_\varphi$ with the gauge choice
can be deduced from (\ref{KFlowRedHam}) and implies 
\begin{equation} \label{NEphiModeqn}
	\(E^\varphi\)^2 = \frac{x^2}{1+1/\delta^2} \(1 +
	2x\frac{N^\prime}{N}\)\,.
\end{equation}
Solving for the Hamiltonian constraint then gives 
\begin{equation}\label{ModEphisolution}
	E^\varphi = \frac{x}{\sqrt{1+1/\delta^2 - c/x}}
\end{equation}
with an integration constant $c$ To simplify notation, we introduce the
parameter $\bar{\delta} = (1 + 1/\delta^2)^{-1}$ from hereon. We solve
for the lapse
\begin{equation}\label{ModNsolution} 
	N(x) = \sqrt{1 -   \frac{c\bar{\delta}}{x}}\,.
\end{equation}
All the consistency conditions, such as preservation of $E^{\varphi}$, still
hold.

The effective line element, with $\beta(K_{\varphi})=-1$ for
$K_{\varphi}=\pi/(2\delta)$, is 
\begin{equation}
	{\rm d} s^2 = \(1 - \frac{c \bar{\delta}}{x}\){\rm d} \tau^2 +
	\frac{\bar{\delta}}{1 - c\bar{\delta}/x} \,{\rm d} x^2
	+x^2{\rm d}\Omega^2\,. 
\end{equation}
A Newtonian limit in Euclidean space would suggest $c\bar{\delta}=2M$, such
that
\begin{equation}\label{ModMetric}
	{\rm d} s^2 = \(1 - \frac{2M}{x}\){\rm d} \tau^2 +
	\frac{\bar{\delta}}{1 - 2M/x} \,{\rm d} x^2
	+x^2{\rm d}\Omega^2\,. 
\end{equation}
(The Newtonian limit is, of course, not a straightforward notion
in Euclidean space. We are merely redefining an integration constant at this
point, but the suggestive identification of $M$ as a potential mass
parameter might be useful.)

The Euclidean solution is valid for $x>2M$. There is no legitimate solution
for $x<2M$ because it would imply two negative components of the metric, but
no such signature is allowed by the effective line elements derived here. The
value $x=2M$ must therefore be interpreted as a boundary of Euclidean space,
rather than a horizon as in the classical Schwarzschild solution. Analyzing
the effective line element for $x$ close to this value then tells us whether
the boundary is regular or forms a conical singularity, just as in
applications of Euclidean gravity to thermodynamic properties of black holes.

\subsection{Black-hole thermodynamics}

Our new stationary solution suggests interesting applications to black-hole
thermodynamics because it automatically appears as a Euclidean effective line
element; no Wick rotation is necessary to derive it.

In (\ref{ModMetric}), the boundary at $x = 2M$ is a conical singularity unless
the metric coefficients $g_{\tau\tau}$ and $g_{xx}$ in an expansion around
this value agree with the 2-dimensional plane in polar coordinates. Defining
$x=2M+\xi$ with small $\xi$, we obtain
\begin{equation}
	1-\frac{2M}{x} \approx \frac{1}{2M}\xi
\end{equation}
and write the 2-dimensional effective line element in $\tau$ and $\xi$ as
\begin{equation}
	{\rm d}s^2 \approx \frac{1}{2M}\xi {\rm d}\tau^2+ 2M\bar{\delta}
	\frac{{\rm d}\xi^2}{\xi}\,.
\end{equation}
Defining $r=\sqrt{8M\xi\bar{\delta}}$ and $\phi=\tau/(4M\sqrt{\bar{\delta}})$
then implies the 2-plane line element ${\rm d}s^2\approx {\rm d}r^2+r^2 {\rm
	d}\phi^2$. We have the full 2-plane rather than a conical singularity
provided that $\phi$ is periodic with period $2\pi$, which implies a period of
$\tau_{\mathrm{T}} = 8\pi M\sqrt{\bar{\delta}}$ for $\tau$.  In black-hole
thermodynamics, this value is usually regarded as the inverse of the Hawking
temperature
\begin{equation} \label{TE}
	T_{\rm E} = \frac{1}{\tau_{\mathrm{T}}}= \frac{1}{8\pi M \sqrt{\bar{\delta}}}
\end{equation}
for our Euclidean black hole.  The corresponding entropy, using the area law,
is
\begin{eqnarray}
	S = \frac{A}{4} = 4 \pi M^{2} = \frac{1}{16 \pi T_{\rm E}^{2} \bar{\delta}}\,,
\end{eqnarray}
and the specific heat
\begin{eqnarray}
	C = \frac{\partial S}{\partial T_{\rm E}} = - \frac{1}{8\pi T_{\rm
			E}^{3}\bar{\delta}} < 0 
\end{eqnarray}
indicates that this black hole is thermodynamically unstable. If $\delta \ll
1$, $\bar{\delta}\approx\delta^2\ll 1$ implies that $T_{\rm E} \gg T_{\rm L}$
where $T_{\rm L}$ is the temperature of a (classical) Lorentzian black hole
with the same mass $M$ as used in (\ref{TE}). The limit of $\delta\to0$ of
such a Euclidean black hole is therefore not a classical black hole, but
rather an object that evaporates very quickly if it is in any way connected to
a Lorentzian space-time.

Properties of higher stationary solutions with $K_{\varphi}=n\pi/(2\delta)$
with integer $n$ can be obtained by substituting $\delta/n$ for $\delta$ in
the preceding equations, or $\bar{\delta}=(1+n^2/\delta^2)^{-1}$ for
$\bar{\delta}$. For small $\delta$, $\bar{\delta}\approx \delta^2/n^2$ is even
smaller than for $n=1$, and evaporation would proceed even more quickly.
We note, however, that a physical interpretation of parameters
such as $M$ and $T$, obtained here exclusively in a Euclidean framework
without the physical basis of a Wick-rotated Lorentzian space-time, is not
clear at this point and needs further study.

\section{Discussion}

We have studied effective line elements derived from anomaly-free spherically
symmetric models of modified canonical gravity. We used canonical effective
methods which are not new, but had not yet been applied in this context. Our
results highlight the importance of off-shell constraint brackets and related
closure conditions. Moreover, the specific form of hypersurface-deformation
brackets in the classical limit is crucial; it cannot be replaced by a
reformulated constrained system (such as a partial Abelianization) even if it
remains first class. These detailed relationships between basic properties of
canonical gravity reveal a direct road from modified structure functions to
coefficients in effective line elements and related geometrical effects. In
particular, if hypersurface-deformation brackets are modified such that the
bracket of two normal deformations changes sign, the corresponding effective
line element changes signature.

Applied to models of loop quantum gravity, therefore, signature change in
effective line elements has been found to be an inevitable consequence if
covariance is preserved by holonomy effects. It is not possible to interpret
changing signs of structure functions as a dynamical instability of matter or
metric perturbations on an otherwise Lorentzian space-time with a well-posed
initial-value problem, as it is sometimes attempted in early-universe
models. Since the only effective line element consistent with modified
hypersurface-deformation brackets becomes Euclidean when holonomy effects are
strong, only 4-dimensional boundary-value problems can be well-posed in this
regime.

We have focused here on spherically symmetric solutions of modified canonical
gravity. New black-hole models with a Euclidean core have been obtained,
correcting the proposal of \cite{BHInt,BHPara} and several more recent follow-up
studies. Our effective line elements also give rise to new non-classical
stationary solutions, but their thermodynamical behavior indicates that they
may not be relevant for a long-term analysis of quantum space-time.

\section*{Acknowledgements}

This work was supported in part by NSF grant PHY-1607414. The work of SB and
DY are supported in part by the Korean Ministry of Education, Science and
Technology, Gyeongsangbuk-Do and Pohang City for Independent Junior Research
Groups at the Asia Pacific Center for Theoretical Physics.

\begin{appendix}
	
	\section{Coordinate transformations and lapse and shift}
	\label{a:LapseShift}
	
	In the main text, we have shown some steps in the derivation of $\delta
	q_{xx}$ from a coordinate transformation of the canonical spherically
	symmetric line element; see (\ref{deltaqxx}). Here, we provide further details
	of a similar derivation of $\delta N^x$ and $\delta N$, given in
	(\ref{NxSphSymm}) and (\ref{NSphSymm}). We will highlight steps in the
	derivation that will change if the hypersurface-deformation brackets are
	deformed.
	
	In order to compute $\delta N^x$, we collect all terms that receive
	differentials ${\rm d}x{\rm d}t$ after substituting $t'=t+\epsilon^0/N$ and
	$x'=x+\epsilon^x-(N^x/N)\epsilon^0$ in the spherically symmetric line element
	\begin{equation} \label{dsSphSymm}
		{\rm d}s^2 = -N(x,t)^2{\rm d}t^2+ q_{xx}(x,t) ({\rm d}x+N^x(x,t){\rm d}t)^2+
		q_{\varphi\varphi}(x,t) ({\rm
			d}\vartheta^2+\sin^2\vartheta{\rm d}\varphi^2)\,.
	\end{equation}
	Such contributions results from all terms except those that have angle
	differentials. In the final expression,
	\begin{eqnarray} \label{deltas1}
		\frac{1}{2}\delta {\rm d}s^2|_{{\rm d}x{\rm d}t} &=& \frac{\epsilon^0}{N}
		\dot{q}_{xx}N^x + \left(\epsilon^x-\frac{N^x}{N}\epsilon^0\right) q_{xx}'N^x+ 
		\frac{\epsilon^0}{N} q_{xx}\dot{N}^x +
		\left(\epsilon^x-\frac{N^x}{N}\epsilon^0\right)  q_{xx}(N^x)'\nonumber\\
		&&+ q_{xx}N^x
		\left(\left(\epsilon^x-\frac{N^x}{N}\epsilon^0\right)'+
		\left(\frac{\epsilon^0}{N}\right)^{\bullet}\right)\nonumber\\
		&&+ q_{xx}
		\left(\epsilon^x-\frac{N^x}{N}\epsilon^0\right)^{\bullet}+ q_{xx}(N^x)^2
		\left(\frac{\epsilon^0}{N}\right)'- N^2 \left(\frac{\epsilon^0}{N}\right)'\,,
	\end{eqnarray}
	the first line results from expanding $q_{xx}$ and $N^x$ in $q_{xx}N^x {\rm
		d}x'{\rm d}t'$, while the second line results from expanding ${\rm d}x'{\rm
		d}t'$ in the same term. The last line contains those terms that produce
	${\rm d}x{\rm d}t$ in an expansion of $q_{xx} {\rm d}x^2$, $q_{xx}(N^x)^2{\rm
		d}t^2$, and $-N^2{\rm d}t^2$, respectively. In an expansion, all these terms
	appear twice, which we have taken into account by dividing by two on the
	left-hand side of the equation.
	
	The first two terms in (\ref{deltas1}) can be recognized as similar
	contributions to $\delta q_{xx}$, given in (\ref{deltaqxx}). The two remaining
	terms in (\ref{deltaqxx}) then change (\ref{deltas1}) to
	\begin{eqnarray} \label{deltas2}
		\frac{1}{2}\delta {\rm d}s^2|_{{\rm d}x{\rm d}t} &=& N^x
		\delta q_{xx} -2q_{xx}N^x\left(\epsilon^x-\frac{N^x}{N}\epsilon^0\right)'-
		2q_{xx}(N^x)^2 
		\left(\frac{\epsilon^0}{N}\right)'\nonumber\\
		&& +
		\frac{\epsilon^0}{N} q_{xx}\dot{N}^x +
		\left(\epsilon^x-\frac{N^x}{N}\epsilon^0\right)  q_{xx}(N^x)'\nonumber\\
		&&+ q_{xx}N^x
		\left(\left(\epsilon^x-\frac{N^x}{N}\epsilon^0\right)'+
		\left(\frac{\epsilon^0}{N}\right)^{\bullet}\right)\nonumber\\
		&&+ q_{xx}
		\left(\epsilon^x-\frac{N^x}{N}\epsilon^0\right)^{\bullet}+ q_{xx}(N^x)^2
		\left(\frac{\epsilon^0}{N}\right)'- N^2 \left(\frac{\epsilon^0}{N}\right)'\,.
	\end{eqnarray}
	Since this result should equal the first-order contribution to $(q+\delta
	q_{xx})(N+\delta N^x)$, the terms on the right other than $N^x\delta q_{xx}$
	give us
	\begin{eqnarray}
		\delta N^x &=& -2N^x\left(\epsilon^x-\frac{N^x}{N}\epsilon^0\right)'-
		2(N^x)^2 
		\left(\frac{\epsilon^0}{N}\right)'\nonumber\\
		&& +
		\frac{\epsilon^0}{N}\dot{N}^x +
		\left(\epsilon^x-\frac{N^x}{N}\epsilon^0\right) (N^x)'+ N^x
		\left(\left(\epsilon^x-\frac{N^x}{N}\epsilon^0\right)'+
		\left(\frac{\epsilon^0}{N}\right)^{\bullet}\right)\nonumber\\
		&&+  \left(\epsilon^x-\frac{N^x}{N}\epsilon^0\right)^{\bullet}+ (N^x)^2
		\left(\frac{\epsilon^0}{N}\right)'- \frac{1}{q_{xx}} N^2
		\left(\frac{\epsilon^0}{N}\right)'\,. 
	\end{eqnarray}
	Notice that we now obtained an inverse metric component in the last term,
	which will be sensitive to modified constraint brackets. If we take all the
	derivatives, the result can be simplified to
	\begin{equation}
		\delta N^x = \dot{\epsilon}^x+ \epsilon^x (N^x)'- N^x(\epsilon^x)'-
		\frac{1}{q_{xx}} \left(N(\epsilon^0)'-\epsilon^0N'\right)\,,
	\end{equation}
	which is (\ref{NxSphSymm}).
	
	The transformation $\delta N$ of the lapse function is derived from an
	expansion of all terms in ${\rm d}s^2$ with at least one ${\rm d}t$. We first
	compute
	\begin{eqnarray}
		\delta {\rm d}s^2|_{{\rm d}t^2} &=& -2N\dot{N} \frac{\epsilon^0}{N}- 2NN'
		\left(\epsilon^x-\frac{N^x}{N}\epsilon^0\right) -2N^2
		\left(\frac{\epsilon^0}{N}\right)^{\bullet}\\
		&& +\left(\dot{q}_{xx}(N^x)^2+ 2q_{xx}N^x\dot{N}^x\right)
		\frac{\epsilon^0}{N}+ \left(q_{xx}'(N^x)^2+ 2q_{xx}N^x(N^x)'\right)
		\left(\epsilon^x-\frac{N^x}{N}\epsilon^0\right)\nonumber\\
		&&+2q_{xx} (N^x)^2 \left(\frac{\epsilon^0}{N}\right)^{\bullet}\nonumber\\
		&&+ 2q_{xx}N^x
		\left(\epsilon^x-\frac{N^x}{N}\epsilon^0\right)^{\bullet}\,,\nonumber 
	\end{eqnarray}
	where the first line results from an expansion of $-N^2{\rm d}t^2$, the next
	two lines from $q_{xx}(N^x)^2{\rm d}t^2$, and the last from $2q_{xx}N^x{\rm
		d}x{\rm d}t$. This result should be equal to the first-order contribution in
	\begin{equation} \label{deltaNqNx}
		-(N+\delta N)^2+ (q_{xx}+\delta q_{xx})(N^x+\delta N^x)^2 =
		-N^2+q_{xx}(N^x)^2 - 2N\delta N+ (N^x)^2\delta q_{xx}+ 2 q_{xx}\delta N^x+
		\cdots
	\end{equation}
	Using $\delta q_{xx}$ in (\ref{deltaqxx}), which does not change for
	holonomy-modified constraints, we can first rewrite
	\begin{eqnarray} \label{deltas3}
		\delta {\rm d}s^2|_{{\rm d}t^2} &=& (N^x)^2\delta q_{xx} -2N\dot{N}
		\frac{\epsilon^0}{N}- 2NN' 
		\left(\epsilon^x-\frac{N^x}{N}\epsilon^0\right) -2N^2
		\left(\frac{\epsilon^0}{N}\right)^{\bullet}\\
		&& +2q_{xx}N^x\dot{N}^x
		\frac{\epsilon^0}{N}+ 2q_{xx}N^x(N^x)'
		\left(\epsilon^x-\frac{N^x}{N}\epsilon^0\right)\nonumber\\
		&&+2q_{xx} (N^x)^2 \left(\left(\frac{\epsilon^0}{N}\right)^{\bullet}-
		N^x\left(\frac{\epsilon^0}{N}\right)'\right) \nonumber\\
		&&+ 2q_{xx}N^x
		\left(\left(\epsilon^x-\frac{N^x}{N}\epsilon^0\right)^{\bullet}-
		N^x\left(\epsilon^x-\frac{N^x}{N}\epsilon^0\right)'\right)  \,,\nonumber 
	\end{eqnarray}
	We next have to collect terms that provide $\delta N^x$ in
	(\ref{deltaNqNx}) before we can read off $\delta N$. We will do so for a
	holonomy-modified expression of the form (\ref{NxSphSymmbeta}), or
	\begin{equation} \label{ShiftGaugebeta}
		\delta N^x = \dot{\epsilon}^x+ \epsilon^x (N^x)'- N^x(\epsilon^x)'-
		\frac{\beta}{q_{xx}} \left(N(\epsilon^0)'-\epsilon^0N'\right)\,,
	\end{equation}
	in order to highlight the lack of cancellations if an unmodified effective
	line element were used. Since the $\delta N^x$-term in (\ref{deltaNqNx}) is
	the only first-order contribution in this expression with a coefficient of
	$q_{xx}$, it must produce all terms in the second, third and fourth line in
	(\ref{deltas3}). By taking derivatives, it is straightforward, if a bit
	tedious, to see that this is indeed the case. However, the $\beta$-dependent
	term in $\delta N^x$ contains the inverse metric component $1/q_{xx}$, which
	produces a term in $q_{xx}\delta N^x$ independent of $q_{xx}$. This remaining
	term should be combined with the first line in (\ref{deltas3}) such that
	$-2N\delta N$ results. We find
	\begin{eqnarray}
		\not\delta N &=& \dot{N} \frac{\epsilon^0}{N} +N' 
		\left(\epsilon^x-\frac{N^x}{N}\epsilon^0\right) +N
		\left(\frac{\epsilon^0}{N}\right)^{\bullet} + \beta \frac{N^x}{N}
		\left(N'\epsilon^0- N(\epsilon^0)'\right) \nonumber\\
		&=& \dot{\epsilon}^0+ N'\epsilon^x + (\beta-1) N^x \frac{N'}{N}\epsilon^0-
		\beta N^x (\epsilon^0)'\,. \label{LapseGaugebeta}
	\end{eqnarray}
	For $\beta=1$, this result is identical with the lapse transformation
	(\ref{LapseGauge}) required for consistency of evolution and gauge. However,
	for $\beta\not=1$, (\ref{LapseGaugebeta}) and (\ref{ShiftGaugebeta}) are not
	consistent with gauge transformations of multipliers of holonomy-modified
	constraints: While (\ref{ShiftGaugebeta}) is implied by a deformed bracket of
	two Hamiltonian constraints, the lapse transformation should not depend on
	$\beta$ because, according to (\ref{deltaNF}) it is sensitive only to the
	unmodified bracket $\{H[N],D[N^a]\}$.
	
	We can obtain consistent results by modifying the original line element
	(\ref{dsSphSymm}) such that expansion coefficients change in the right
	way. This modification then implies the effective line element. We first note
	that all contributions to (\ref{LapseGaugebeta}) other than the
	$\beta$-dependent term originate from an expansion of $-N^2{\rm d}t^2$. If we
	modify this contribution to the line element to $-\beta N^2{\rm d}t^2$, all
	terms in the first line of (\ref{LapseGaugebeta}) are proportional to $\beta$,
	and the classical cancellations are revived:
	\begin{eqnarray}
		\delta_{\beta} N &=& \beta \dot{N} \frac{\epsilon^0}{N} +\beta N' 
		\left(\epsilon^x-\frac{N^x}{N}\epsilon^0\right) +\beta N
		\left(\frac{\epsilon^0}{N}\right)^{\bullet} + \beta \frac{N^x}{N}
		\left(N'\epsilon^0- N(\epsilon^0)'\right) \nonumber\\
		&=& \beta\left(\dot{\epsilon}^0+ N'\epsilon^x - N^x (\epsilon^0)'\right)=
		\beta \delta N\,,
	\end{eqnarray}
	where $\delta N$ is the classically expected transformation. If $\beta$
	depends on $t$ or $x$, the expansion of $-\beta N^2$ will recieve additional
	terms, which can be combined to
	\begin{eqnarray} \label{Absorb}
		&&-2\beta N\delta N- \dot{\beta} \epsilon^0 N- \beta'
		\left(\epsilon^x-\frac{N^x}{N}\epsilon^0\right) N^2\nonumber\\
		&=& -2N\beta\left( \delta N \frac{1}{2}\frac{\dot{\beta}}{\beta} \epsilon^0+
		\frac{1}{2}\frac{\beta'}{\beta} (N\epsilon^x- N^x\epsilon^0)\right)\nonumber\\
		&=& -2 \sqrt{\beta}N \left((\sqrt{\beta}\epsilon^0)^{\bullet}+
		(\sqrt{\beta}N)'- N^x(\sqrt{\beta}\epsilon^0)'\right)\,.
	\end{eqnarray}
	Gauge transformations of lapse and shift for holonomy-modified constraints are
	therefore consistent with the effective line element (\ref{dsEffective}), but
	not with the classical form of the line element. Equivalently, as shown by
	(\ref{Absorb}), one can work with the usual line element after a redefinition
	of the lapse function $N$ and gauge parameter $\epsilon^0$ absorbing
	$\sqrt{\beta}$, provided $\beta>0$. If $\beta<0$, one can absorb
	$\sqrt{-\beta}$ in the lapse function and gauge parameter of Euclidean space.
	
	\section{Gauge transformation of the spatial metric}
	\label{a:q}
	
	The spatial metric has two independent components in spherically symmetric
	models, $q_{xx}$ and $q_{\varphi\varphi}$. In this appendix, we show that
	their gauge transformations, expressed in terms of their time and space
	derivatives, are not modified for non-classical $f_1$ and $f_2$. The result
	for $q_{xx}$ is important in our derivation of lapse and shift
	transformations, but the example of $q_{\varphi\varphi}$ is simpler. We
	present it first as a warm-up.
	
	Since $q_{\varphi\varphi}=E^x$, we can directly use (\ref{deltaEx}) and write
	\begin{equation}
		\delta q_{\varphi\varphi} = \delta E^x = 2 \epsilon^0 \sqrt{E^x}
		f_2(K_{\varphi}) + \epsilon^x (E^x)'\,.
	\end{equation}
	This expression is modified for non-classical $f_2$. However, if we use the
	modified equation of motion (\ref{Exdot}), solved for
	\begin{equation}
		2\sqrt{E^x} f_2(K_{\varphi}) = \frac{1}{N}\left(\dot{E}^x-N^x(E^x)'\right)\,,
	\end{equation}
	we obtain
	\begin{eqnarray*}
		\delta q_{\varphi\varphi} &=& \frac{\epsilon^0}{N}
		\left(\dot{E}^x-N^x(E^x)'\right)+ \epsilon^x(E^x)'\\
		&=& \frac{\epsilon^0}{N} \dot{q}_{\varphi\varphi}+
		\left(\epsilon^x-\frac{N^x}{N} \epsilon^0\right) q_{\varphi\varphi}'\,.
	\end{eqnarray*}
	There are no modifications in this equation.
	
	For $q_{xx}$, we proceed in a similar way and write
	\begin{eqnarray*}
		\delta q_{xx} &=& \delta\left(\frac{(E^{\varphi})^2}{E^x}\right) =
		-\frac{(E^{\varphi})^2}{(E^x)^2} \delta E^x+ 2\frac{E^{\varphi}}{E^x} \delta
		E^{\varphi}\\
		&=& -\frac{(E^{\varphi})^2}{(E^x)^2} \left(2 \epsilon^0 \sqrt{E^x} \;
		f_2(K_{\varphi}) + \epsilon^x 
		(E^x)'\right)\\
		&&+ 2\frac{E^{\varphi}}{E^x} \left(\epsilon^0 \sqrt{E^x} K_x
		\frac{{\rm d}f_2(K_{\varphi})}{{\rm 
				d}K_{\varphi}} +  \frac{\epsilon^0 E^{\varphi}}{2\sqrt{E^x}} \frac{{\rm
				d}f_1(K_{\varphi})}{{\rm d}K_{\varphi}} + (\epsilon^x
		E^{\varphi})'\right)\\
		&=& 2 \frac{E^{\varphi}}{\sqrt{E^x}} K_x \frac{{\rm d}f_2(K_{\varphi})}{{\rm 
				d}K_{\varphi}} \epsilon^0+
		\left(\frac{\epsilon^x(E^{\varphi})^2}{E^x}\right)'+
		\frac{(E^{\varphi})^2}{E^x}(\epsilon^x)'+ \frac{(E^{\varphi})^2}{(E^x)^{3/2}}
		\epsilon^0\left(\frac{{\rm d}f_1(K_{\varphi})}{{\rm  
				d}K_{\varphi}}- 2f_2(K_{\varphi}) \right)\,.
	\end{eqnarray*}
	If we now use the equations of motion to eliminate $K_x$, we have
	\begin{eqnarray*}
		\delta q_{xx} &=& \frac{2\epsilon^0E^{\varphi}}{E^x N} \left(\dot{E}^{\varphi}-
		\frac{E^{\varphi}}{2E^x} \left(\dot{E}^x+ N\sqrt{E^x} \left(\frac{{\rm
				d}f_1(K_{\varphi})}{{\rm d}K_{\varphi}}-2f_2(K_{\varphi})\right)
		-N^x(E^x)'\right)  - (N^xE^{\varphi})'\right)\\
		&&+
		\left(\frac{\epsilon^x(E^{\varphi})^2}{E^x}\right)'+
		\frac{(E^{\varphi})^2}{E^x}(\epsilon^x)'+ \frac{(E^{\varphi})^2}{(E^x)^{3/2}} 
		\epsilon^0\left(\frac{{\rm d}f_1(K_{\varphi})}{{\rm  
				d}K_{\varphi}}- 2f_2(K_{\varphi}) \right)\\
		&=& \frac{\epsilon^0}{N} \left(\frac{(E^{\varphi})^2}{E^x}\right)^{\bullet}+
		\left(\epsilon^x-\frac{N^x}{N}\epsilon^0\right)
		\left(\frac{(E^{\varphi})^2}{E^x}\right)'- 2\frac{(E^{\varphi})^2}{E^x}
		\frac{\epsilon^0}{N} (N^x)'+ 2\frac{(E^{\varphi})^2}{E^x} (\epsilon^x)'\\
		&=& \frac{\epsilon^0}{N} \dot{q}_{xx}+
		\left(\epsilon^x-\frac{N^x}{N}\epsilon^0\right)
		q_{xx}'- 2q_{xx} \frac{\epsilon^0}{N} (N^x)'+ 2q_{xx} (\epsilon^x)'\,.
	\end{eqnarray*}
	Also this equation is unmodified by holonomy terms. Note that the condition
	(\ref{f1f2}) for anomaly-free constraints would simplify the derivation, but
	it is not required for $\delta q_{xx}$, expressed in terms of its time and
	space derivatives, to be of the classical form.
	
\end{appendix}

\end{document}